\documentclass[12pt]{article}

\usepackage{amsmath,amssymb}
\usepackage{graphicx}% Include figure files
\usepackage{caption}
\usepackage{color}% Include colors for document elements
\usepackage{dcolumn}% Align table columns on decimal point
\usepackage{bm}% bold math
\usepackage[numbers,super,comma,sort&compress]{natbib}
\usepackage{multirow}
\usepackage{textgreek}
\usepackage{epstopdf}
\usepackage{siunitx}
\usepackage{hyperref}
\usepackage{dcolumn}
\usepackage{tikz}
\usepackage{slashbox}

\newcolumntype{.}{D{.}{.}{-1}}

\definecolor{background-color}{gray}{0.98}

\title{Depurated Inversion Method for Orbital--Specific Exchange 
Potentials}
\author{A.M.P. Mendez\footnote{alemendez@iafe.uba.ar},
D.M. Mitnik, J.E. Miraglia 
\thanks{Instituto de Astronom\'{\i}a y F\'{\i}sica del Espacio, 
CONICET--UBA, Buenos Aires, Argentina.}}

\begin{document}

\maketitle

\begin{abstract}
This work presents exchange potentials for specific orbitals 
calculated by inverting Hartree--Fock wavefunctions. This was 
achieved by using a Depurated Inversion Method. 
The basic idea of the method relies upon the 
substitution of Hartree--Fock orbitals and eigenvalues into the 
Kohn--Sham equation. Through inversion, the corresponding effective 
potentials were obtained. Further treatment of the inverted 
potential should be carried on. The depuration is a careful 
optimization which eliminates the poles and also ensures the 
fullfilment of the appropriate boundary conditions. The procedure 
developed here is not restricted to the ground state or to a 
nodeless orbital and is applicable to all kinds of atoms. As an 
example, exchange potentials for noble gases and term--dependent 
orbitals of the lower configuration of Nitrogen are calculated. 
The method allows to reproduce the input energies and wavefunctions 
with a remarkable degree of accuracy.
\end{abstract}

\clearpage

%%%%%%%%%%%%%%%%%%%%%%%%%%%%%%%%%%%%%%%%%%%%%%%%%%%%%%%%%%%%%%%%%%%%%
%%%%%%%%%%%%%%%%%%%%%%%%%%%%%%%%%%%%%%%%%%%%%%%%%%%%%%%%%%%%%%%%%%%%%
\section{\sffamily \Large INTRODUCTION} 
\label{sec:introduction}

The successful idea of replacing a many--body, non--local 
interaction by an effective one--electron equation 
opened up the possibility of studying extremely complex systems 
with high accuracy.
In this context, the success of the Kohn--Sham density--functional 
theory\cite{HohenberKohn:64,KohnSham:65} (DFT) began when crucial 
developments in its exchange--correlation 
terms gave the theory predictive power to compete with 
well--developed wavefunction methods\cite{Becke:14}.
The importance of the exchange--correlation potentials in chemical 
physics has been emphasized by Bartlett\cite{Bartlett:10,Verma:12}.
Exchange potentials are in general constructed by local 
approximations to the nonlocal Hartree--Fock exchange operator
(i.e. the Slater potential\cite{Slater:51}, the optimized 
effective potential\cite{Sharp:53,Talman:76,Talman:89}, 
the Krieger--Li--Iafrate\cite{Krieger:92} and several 
others\cite{Gorling:92,Yang:02,Staroverov:06,Ryabinkin:13}). 

The atomic collision community, on the other hand, is also eager to 
accurately determine effective one--electron local potentials 
which would allow to generate in a simpler way the wavefunctions of the 
particles interacting in a scattering process. 
In particular, we need to represent an orthonormal set of bound 
and continuum states to calculate the transition probabilities. 
This should include detailed $nl$--orbital potentials, 
a feature missing in most of the standard density functional
methods. 
Soft pseudopotentials like {\sc abinit}\cite{abinit} or 
{\sc uspp}\cite{Vanderbilt} cannot be used because they 
overlook the information of the internal region of the wavefunctions. 
The features of this region can play a very important role, 
such as the cusp conditions in the processes of electron capture and 
ionization. 
In an attempt to meet the needs of both chemist and collisionist 
communities, we strove to obtain accurate and simple specific 
$nl$--orbital local potentials. 

How to determine central potentials from known electron 
wavefunctions and densities is a well studied subject in the DFT 
community\cite{Wu:03,Gaiduk:13,Ryabinkin:15}.
The extraction of the true Kohn--Sham exchange--correlation 
potential from near--exact electronic densities has been 
demonstrated, with particular reference to two--electron systems 
like He\cite{Mura:97}, He--isoelectronic ions\cite{Umrigar:94}, 
and H$_2$\cite{Gritsenko:97,Mura:97} as well as exact
soluble models (for example, an external harmonic potential as in 
Filippi \textit{et al}\cite{Filippi:94}).

Some other works start with a particular Kohn--Sham potential and
solve the corresponding equations, obtaining the KS 
orbitals\cite{Schipper:97,deSilva:12,Kananenka:13}. 
Through inversion, they obtain a reconstructed KS potential, which 
agrees almost everywhere with the original one, except in 
some regions where huge oscillations arise. 
In some cases, the reconstructed potential may be distorted beyond
recognition\cite{Mura:97,Jacob:11}.
The same type of procedure was suggested many years ago 
by Hilton \textit{et al.}, 
in applications circumscribed to the calculation of 
photoionization processes of atoms\cite{Hilton:77,Suzer:77},
water\cite{Hilton:79} and other molecules\cite{Hilton:80,Crljen:87}.
These papers, in turn, refer to the earlier work in atomic 
polarizability carried out by Sternheimer\cite{Sternheimer:54} 
and Dalgarno and Parkinson\cite{Dalgarno:59}. 
However, they focused on the final photoionization cross section 
results, and did not provide details about the quality of the 
potentials and the wavefunctions they generated.

Assuming the validity of the separation between exchange 
and correlation functionals, we will focus here only on the 
calculation of the exchange contribution to the potential.
Since Hartree--Fock does not include the correlations, 
our approach allows to obtain the ``exact''
one--electron local potential representing the exchange interactions. 
Strictly speaking, the method does not rely on the KS inversion formula 
since the Hartree--Fock solutions were the ones used for the inversion.
That is, we solved a KS--type equation, but rather than having 
KS--orbitals, we operated directly with the Hartree--Fock 
wavefunctions.
For open--shell atoms, we were able to find orbital 
spin--polarized exchange potentials, this being crucial, for 
instance, to find the hyperfine coupling 
constants\cite{Barone:94,Kaupp:10}.

However, this is not a simple task, and probably that is
why the method presented here has not been widely applied in the past. 
If the wavefunction has nodes, it will produce huge poles in the 
potential. 
Moreover, even for nodeless states, the asymptotic decaying behavior 
of the bound wavefunctions produces severe numerical difficulties, 
making the inversion operation intractable sometimes.
In our method, a depuration procedure follows the inversion. 
This depuration implies, first, the annihilation of the poles.
Then, a careful optimization of the potential which 
ensures the fulfillment of the appropriate boundary conditions. 

The work is organized as follows. 
Section \ref{sec:theory} describes the method, which includes 
the inversion procedure (\ref{subsec:directinversion}), 
the potential depuration (\ref{subsec:depuration}) and its further 
optimization (\ref{subsec:optimization}).
Section \ref{sec:results} presents the resulting effective potentials 
for the orbitals corresponding to the ground states of different 
noble gases, including a thorough 
examination of the wavefunctions generated by these potentials 
(\ref{subsec:resultsDIM}).
The corresponding exchange potentials are discussed in 
(\ref{subsec:exchange}), comparing the potentials for 
specific--$nl$ orbitals with averaged potentials.
Results of the same calculations for the Nitrogen atom are provided in
(\ref{subsec:Nitrogen}).
Atomic units are used unless otherwise specified.

%%%%%%%%%%%%%%%%%%%%%%%%%%%%%%%%%%%%%%%%%%%%%%%%%%%%%%%%%%%%%%%%%%%%%
%%%%%%%%%%%%%%%%%%%%%%%%%%%%%%%%%%%%%%%%%%%%%%%%%%%%%%%%%%%%%%%%%%%%%
\section{\sffamily \Large THEORY}
\label{sec:theory}

%%%%%%%%%%%%%%%%%%%%%%%%%%%%%%%%%%%%%%%%%%%%%%%%%%%%%%%%%%%%%%%%%%%%%
\subsection{\sffamily \Large The direct inversion method}
\label{subsec:directinversion}

The radial part of the Schr\"{o}dinger equation for an electron in a 
local and central potential~is 
\begin{equation}
\left[ 
-\frac{1}{2}\frac{d^{2}}{dr^{2}} + \frac{l(l+1)}{2r^{2}}+V_{nl}(r) 
\right] u_{nl}(r) = 
\varepsilon _{nl}\,u_{nl}(r) \, .  
\label{eq:SchrodKS}
\end{equation}
We assume the following hypothesis:  If the wavefunctions $u_{nl}$ 
are replaced by the solutions of an Hartree--Fock calculation 
$u_{nl}^{\mathrm{HF}}$, then, the corresponding effective local 
potentials $V_{nl}^{\mathrm{HF}}$ that generate such wavefunctions
should exist. 
Based on this we converted the HF method 
into a set of Kohn--Sham equations, whose solutions are the 
Hartree--Fock wavefunctions:
\begin{equation}
\left[ 
-\frac{1}{2}\frac{d^{2}}{dr^{2}} + \frac{l(l+1)}{2r^{2}} + 
V_{nl}^{\mathrm{HF}}(r) 
\right] u_{nl}^{\mathrm{HF}}(r)
   = \varepsilon_{nl}^{\mathrm{HF}}\, u_{nl}^{\mathrm{HF}}(r) \, .
\label{eq:KS}
\end{equation}
The effective potentials given by, 
\begin{equation}
V_{nl}^{\mathrm{HF}}(r) = V^{\mathrm{C}}(r) + 
V^{\mathrm{dir}}(r) + V_{nl}^{\mathrm{x}}(r) \, ,  
\label{eq:veff}
\end{equation}
are composed of the external potential $V^{\mathrm{C}}$ 
(the Coulomb field of the nucleus), the direct (or Hartree)
potential $V^{\mathrm{dir}}$ (the electrostatic electron repulsion), 
and the orbital exchange potentials $V_{nl}^{\mathrm{x}}$. 
We have ignored the correlation term since the HF solutions
do not include~it. 

Since the solutions $u_{nl}^{\mathrm{HF}}$ are known 
(calculated numerically with the {\sc hf} code by C. F. Fischer 
\cite{Fischer:97}, and the {\sc nrhf} code by W. Johnson 
\cite{Johnson:07}) we proceeded to directly invert
the Kohn--Sham--type equations:
\begin{equation}
V_{nl}^{\mathrm{HF}}(r) = 
\frac{1}{2}\frac{1}{u_{nl}^{\mathrm{HF}}(r)}
\frac{d^2}{dr^{2}}u_{nl}^{\mathrm{HF}}(r) - 
\frac{l(l+1)}{2r^{2}}+\varepsilon _{nl}^{\mathrm{HF}} \, ,
\label{eq:V}
\end{equation}
obtaining the \emph{HF inverted potential} 
$V_{nl}^{\mathrm{HF}}(r)$.
Assuming a Coulombic--type shape, it is convenient to define an 
\emph{HF inverted charge}
\begin{equation}
Z_{nl}^{\mathrm{HF}}(r) \equiv -r \, V_{nl}^{\mathrm{HF}}(r) \, .
\label{eq:Zeff}
\end{equation}

The direct computation of (\ref{eq:V}) is known to pose
serious numerical problems\cite{Mura:97}.
\textit{First}, the presence of (genuine) nodes in the wave function 
to be inverted produces poles and unrealistic features around them. 
This has led to the general consensus that the inversion method can 
only be used for nodeless orbitals\cite{Kananenka:13}. 
\textit{Second}, numerical rounding up of the exponential 
decay of the bound states hinders the corresponding inverted 
potential from having the physically desired asymptotic form. 
Moreover, there is a \textit{third} problem at the very heart 
of the Hartree Fock method: the exact solutions may have oscillations 
(and therefore, spurious nodes) in the large--r or ``tail'' region of 
the functions. 
The existence of these spurious nodes in Hartree Fock was already 
suggested by Fischer\cite{Fischer:97}. 
This failure is not caused by the numerical scheme but it is 
inherent to the method. Probably, these nodes are surviving 
long--range exchange effects due to the non--local character of the 
Hartree--Fock wavefunctions: the behavior of a particular orbital 
depends on all others. 
We have found the same spurious nodes at the same places even  
using different numerical codes.
As a general rule, the spurious nodes appear at very long distances, 
in regions where the amplitude of the wavefunction is very small. 
Therefore, their existence has no practical consequences, and they 
can be ignored in any general Hartree--Fock calculation.  
However, this is not true as far as the inversion procedure is 
concerned, as we will discuss in the next section.
Other examples where the presence of orbital nodes (both formal and 
those in the tail region) can be problematic in inversion procedures 
can be found in the literature  (see for instance Peach 
{\it et al.}\cite{Peach:12}).

%%%%%%%%%%%%%%%%%%%%%%%%%%%%%%%%%%%%%%%%%%%%%%%%%%%%%%%%%%%%%%%%%%%%%
\subsection{\sffamily \Large The depurated inversion method}
\label{subsec:depuration}

The difficulties mentioned above make it very hard to obtain the correct 
$V_{nl}^{\mathrm{HF}}(r)$ potentials using the simple inversion formula
given by Eq. (\ref{eq:V}). 
To overcome these troubles we have developed a depurated inversion 
method (DIM) which optimizes the effective charges 
rather than the effective potentials. 
We managed to constrain any potential to have the right
boundary conditions by enforcing the {\it effective depurated inverted 
charge} to behave as follows:
\begin{equation}
Z_{nl}^{\mathrm{DIM}}(r) \, \rightarrow 
\left\{ 
\begin{array}{ll}
Z_{N}  \ \  & \ \ \text{as\ \ }r  \rightarrow 0\  \\ 
1           & \ \ \text{as\ \ }r  \rightarrow \infty \ 
\end{array}
\right.  
\label{eq:Zasympt}
\end{equation}
where $Z_N$ is the nuclear charge.
Once the charge is determined at the boundaries, we can obtain 
a smooth analytic expression for $Z_{nl}^{\mathrm{DIM}}(r)$, fitting the 
$Z_{nl}^{\mathrm{HF}}(r)$ for the largest possible range, 
except in the neighborhood of the nodes.
All this can be accomplished by imposing the 
effective DIM charge to fit the following 
analytical expression:
\begin{equation}
Z_{nl}^{\mathrm{DIM}}(r)= \sum_{j}\alpha _{j}e^{-\beta _{j}r}+1 \, ,  
\label{eq:atomseq}
\end{equation}
with $\Sigma _{j}\alpha _{j}=Z_{N}-1$.

\begin{figure}
\centering
\includegraphics[width=3.33in]{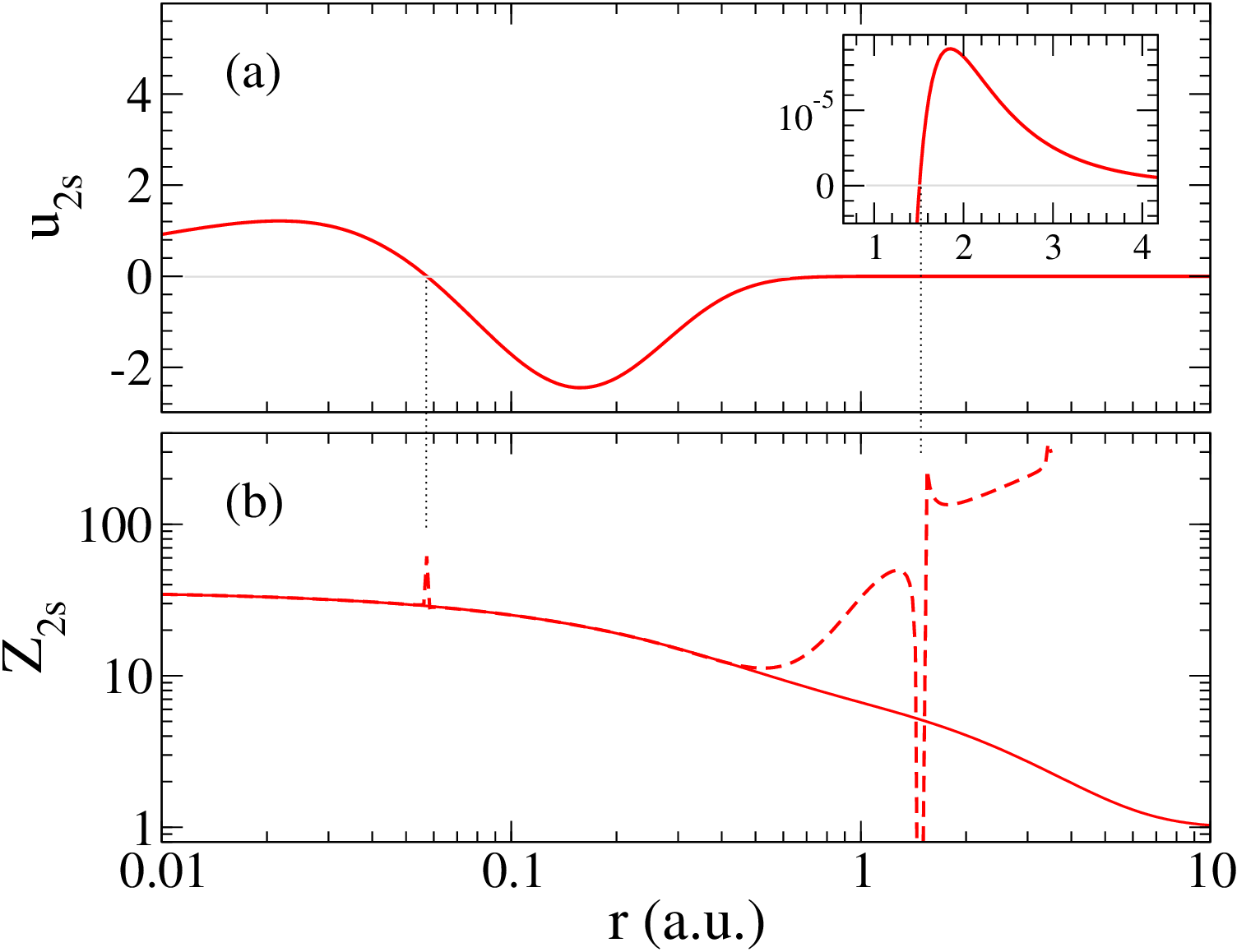} 
\caption{(a) Hartree--Fock orbital 
$u_{2s}^{\mathrm{HF}}$ corresponding to the ground state of the Kr 
atom. It presents two nodes, a genuine one at $r \approx 0.06$ a.u., 
and a spurious one at $1.51$ a.u. (shown in the inset).
(b) Dashed line: The corresponding inverted effective charge 
$Z_{2s}^{\mathrm{HF}}(r)$, spoiled by the presence of poles.
Solid line: Depurated $Z_{2s}^{\mathrm{DIM}}(r)$ effective charge.}
\label{fig:2sKr}
\end{figure}

As a clear instance of the numerical problems mentioned and the way 
propose to solve them, we show, in  Figure \ref{fig:2sKr}, the 
orbital $u_{2s}^{\mathrm{HF}}(r)$ of the ground 
state of the Kr atom (part (a)), and its correspondent effective 
charge $Z_{2s}^{\mathrm{HF}}(r)$ (dashed line curve, in part (b)). 
First, note that the $2s$ orbital has a genuine node at 
$r \approx 0.06$ a.u.  which produces the first pole in the effective 
charge, as shown in the lower graph.
The node appears at a relatively low--$r$ value, so the 
corresponding charge (see Eq. (\ref{eq:Zeff})) is not very sensitive 
to its presence. Therefore, it is very easy to eliminate the pole 
from the effective charge (by just erasing a few points around this 
radius).

All the bound wavefunctions decay exponentially beyond the last 
turning point $r_{tp}$, defined as the position in 
which the energy equals the effective potential. 
At first glance, it seems that the turning point of $u_{2s}(r)$ 
is located around $r_{tp} \approx 0.25$, and from that point on,  
the wavefunction should start to decay exponentially.
From the numerically point of view, $r \approx 10 \, r_{tp}$ is 
a good point to stop the inversion, since beyond there, the effective 
charge could begin to diverge. 
Thus, one might infer that by erasing the points belonging 
to the neighborhood of the first node, and by stopping the inversion 
about $10 \, r_{tp}$, the inversion procedure 
will work well.
However, the dashed curve in Fig.~\ref{fig:2sKr}(b) shows a 
completely unphysical $Z^{\mathrm{HF}}_{2s}(r)$ resulting from the 
inversion.
A very careful examination of the $u_{2s}^{\mathrm{HF}}(r)$ orbital 
function evidences the presence of a spurious node at $r=1.51$ a.u., 
in a region where the amplitude of the wavefunction is 
less than $10^{-4}$ times the maximum value (see the 
inset of Fig. \ref{fig:2sKr}(a)).
Even though this node is completely innocuous for practical matters,  
it produces devastating effects in the inversion procedure, 
evidenced by the second huge peak in the 
$Z_{2s}^{\mathrm{HF}}(r)$ curve (see Figure \ref{fig:2sKr}(b)). 
This pole is so big that it affects a 
broad vicinity and causes the abrupt rising of the effective charge 
for $r>0.5$ a.u.. This is really a surprising result since 
{\it a priory} there is no reason to suspect that a negligible 
oscillation in the tail of the wavefunction would produce such a 
big drawback at small distances.
Care must be taken then to discard these kind of undesired effects.

%%%%%%%%%%%%%%%%%%%%%%%%%%%%%%%%%%%%%%%%%%%%%%%%%%%%%%%%%%%%%%%%%%%%%
\subsection{\sffamily \Large Optimization}
\label{subsec:optimization}

The adjustment of the parameters $\alpha _{j}$ and $\beta _{j}$ 
also requires carefull work.
The key issue in the successful approximation is the region 
chosen for the fitting: it has to be 
as large as possible, in such a way that $Z_{nl}^{\mathrm{DIM}}(r)$ 
overlaps the inverted $Z_{nl}^{\mathrm{HF}}(r)$ across a broad 
range, allowing an accurate fitting procedure, but discarding the 
points surrounding the nodes. 
Also, the inversion must be halted at a particular 
(as large as possible) $r$ value, as soon as the amplitude of the 
function is too small. Further on, the inversion procedure 
diverges.
Another issue to consider is the self consistency within 
the computer codes used in the calculations and the particular  
code used to generate the input wavefunctions. 
To that end, we make sure that the same specific 
numerical grid is used, including the derivatives and integrals 
at the same pivots.
The optimization procedure is completed by a number of iteration 
steps, in which the parameters are optimized to give accurate 
energies and wavefuntions.

Most density functional approximation methods are based on 
a variational principle, minimizing the density functionals 
according to energy (others are defined by density). 
Without underestimating its importance, energy is only
one of the many parameters that characterizes a quantum state. 
Different trial functions (having different forms) can produce, 
through a variational procedure, the same final energy. 
A simple example is given by Bartschat \cite{Albright:93,Bartschat:96} 
in which two different potentials (one having exchange, the other 
omitting it) led to producing very similar and accurate energies 
of the Rydberg series in several quasi--one electron systems. 
However, a further examination of these potentials shows large 
discrepancies in scattering calculations \cite{BartschatBray:96}.
Therefore, in addition to the energy criterion, we have included 
in our optimization method a variational procedure to reproduce 
accurately the wavefunctions. This is achieved by
optimizing the mean values $\langle 1/r \rangle$ (which 
characterize the quality of the wavefunction near the origin), and 
$\langle r \rangle$ (probing it at longer distances). 
Furthermore, we defined the quantity
\begin{equation}
 \delta = 1-\frac{\int{u_{nl}^{\mathrm{HF}}(r) u_{nl}^{\mathrm{DIM}}(r)\,
 dr}}{ \int{ \rho_{nl}^{\mathrm{HF}}(r) dr}}\,.
 \label{eq:orthogonality}
\end{equation}
to determine the accuracy of the orbitals generated by the 
diagonalization of the DIM potentials and the original HF orbitals. 

The effective depurated inversion charge $Z_{2s}^{\mathrm{DIM}}(r)$ 
corresponding to the $2s$ orbital of the Kr atom resulting from 
the optimization is shown --solid curve-- in Figure \ref{fig:2sKr}(b).
As seen in the figure, both boundary conditions are fulfilled 
(at the origin, $Z_{2s} \rightarrow 36$, and asymptotically 
$Z_{2s} \rightarrow 1$, as stated in Eq. (\ref{eq:Zasympt})).

%%%%%%%%%%%%%%%%%%%%%%%%%%%%%%%%%%%%%%%%%%%%%%%%%%%%%%%%%%%%%%%%%
\section{\sffamily \Large RESULTS}
\label{sec:results}

\subsection{\sffamily \Large DIM Potentials, energies and mean values}
\label{subsec:resultsDIM}

The fitting parameters $\alpha _{j}$ and $\beta _{j}$ defining 
the effective charges $Z_{nl}^{\mathrm{DIM}}(r)$ in 
Eq. (\ref{eq:atomseq}) for the noble 
gases Helium, Neon, Argon and Krypton, are given in 
Table \ref{tab:parameters}. 
We have limited the $\alpha _{j}$ and $\beta _{j}$ to six 
(about two per shell).
For Kr, we would probably need two more since there 
are four shells involved. 
\begin{table}
\caption{Fitting parameters for the effective charge 
$Z_{nl}^{\mathrm{DIM}}(r)$ for He, Ne, Ar, and Kr, 
applying Eq. (\ref{eq:atomseq}).}
\label{tab:parameters}
\centering
\begin{tabular}{cccccccc}
\hline
   & $nl$ & $\alpha$ & $\beta$   &    & $nl$ & $\alpha$ & $\beta$ \\
\hline
He & $1s$ & -0.31745 & 5.04372   & Kr & $1s$ & 5.49263  & 0.884768  \\
   &      & 1.31745  & 2.50032   &    &      & 3.94437  & 16.8769   \\
 \vspace*{1ex}
   &      &    -     &   -       &    &      & 25.5630  & 3.10032   \\
Ne & $1s$ & 7.367687 & 2.417275  &    & $2s$ & 9.63120  & 0.575832  \\
   &      & 1.300360 & 0.126396  &    &      & 1.84650  & 25.53280  \\
 \vspace*{0.09cm}
   &      & 0.331953 & 13.15820  &    &      & 23.5223  & 4.543350  \\
   & $2s$ & 0.297739 & 17.99390  &    & $2p$ & 3.20530  & 20.83535  \\
   &      & 0.668081 & 0.067288  &    &      & 23.6172  & 3.928520  \\
 \vspace*{0.09cm}
   &      & 8.03418  & 2.47221   &    &      & 8.17750  & 0.636486  \\
   & $2p$ & 1.353049 & 8.56948   &    & $3s$ & 6.52203  & 0.547357  \\
   &      & 0.335881 & 0.464942  &    &      & 24.4475  & 3.657030  \\
 \vspace*{1ex}
   &      & 7.311070 & 2.090634  &    &      & 4.03047  & 16.61770  \\
Ar & $1s$ & 6.727570 & 6.177720  &    & $3p$ & 23.13135 & 4.010523  \\
   &      & 4.751090 & 1.343560  &    &      & 3.325360 & 20.41890  \\
 \vspace*{0.09cm}
   &      & 5.521340 & 0.859981  &    &      & 8.543290 & 0.821218  \\
   & $2s$ & 8.90271  & 1.09779   &    & $3d$ & 10.05320 & 1.04843   \\ 
   &      & 2.36850  & 2.93144   &    &      & 21.81544 & 4.25746   \\ 
 \vspace*{0.09cm}
   &      & 5.72879  & 6.95913   &    &      & 3.131360 & 20.6087   \\
   & $2p$ & 4.96956  & 6.14455   &    & $4s$ & 3.65988  & 0.49000   \\ 
   &      & 1.48464  & 10.86843  &    &      & 26.4565  & 3.17799   \\
 \vspace*{0.09cm}
   &      & 10.5458  & 1.30005   &    &      & 4.88362  & 15.2031   \\
   & $3s$ & 10.3202  & 2.33169   &    & $4p$ & 7.35713  & 1.00142   \\ 
   &      & 4.27115  & 7.33678   &    &      & 24.2321  & 3.7309    \\  
 \vspace*{0.09cm}
   &      & 2.40865  & 0.407463  &    &      & 3.41077  & 22.5680   \\
   & $3p$ & 8.43753  & 3.49259   &&&&\\
   &      & 2.18200  & 10.8595   &&&&\\
   &      & 6.38047  & 1.07080   &&&&\\
\hline
\end{tabular}
\end{table}
Having these effective charges, we built the corresponding DIM 
potentials $V_{nl}^{\mathrm{DIM}}(r)$. 
By solving the Schr\"odinger equation (Eq. (\ref{eq:SchrodKS})), 
we obtained the solutions $u_{nl}^{\mathrm{DIM}}(r)$ and the 
corresponding energies $\varepsilon _{nl}^{\mathrm{DIM}}$.
The comparison between the results obtained from the diagonalization 
of the Hamiltonian with the $V_{nl}^{\mathrm{DIM}}(r)$ effective 
potential and the original Hartree--Fock orbitals are presented in 
Table \ref{tab:results}.
It is remarkable that with such simple analytical expressions for the 
potentials we were able to reproduce exactly the same energies as the
HF method. The only exception is the $4p$ orbital 
of Kr, in which both calculations agree up to the fifth significant 
figure. 
The fitting procedure also allows to reproduce the original HF 
wavefunctions with an outstanding degree of accuracy. 
The agreement between the HF orbitals
$u_{nl}^{\mathrm{HF}}(r)$ and the solutions $u_{nl}^{\mathrm{DIM}}(r)$ 
can be tested through the comparison of the mean values 
$\langle r \rangle$ and $\langle 1/r \rangle$, and the computation 
of quantity $\delta$ defined by Eq. (\ref{eq:orthogonality}).
The mean values agree in about $0.1\%$ while the values of $\delta$ 
are about $10^{-5}$.
\begin{table}
\caption{Total and orbital energies, mean and $\delta$ values for He, Ne, Ar and 
Kr atoms obtained from DIM effective potentials (upper rows) compared 
with the original Hartree--Fock values (lower rows).} 
\label{tab:results}
\centering
\begin{tabular}{c *{1}{S[table-format=6.4]}|c *{1}{S[table-format=5.6]}
*{1}{S[table-format=2.6]} *{1}{S[table-format=2.6]} c}
\firsthline
   & {$E$} & {$nl$} &  {$\epsilon$} & {$\langle r \rangle$}
   & {$\langle 1/r \rangle$} & {$\delta$} \\ 
\hline
He & -2.8616 & $1s$ & -0.917956   & 0.927313 & 1.687251 & $8\times10^{-10}$ \\
   & -2.8617 &      & -0.917956   & 0.927273 & 1.687282 &   \\
Ne & -128.4978 & $1s$ & -32.772447  & 0.157491 & 9.621450 & $2\times10^{-6}$  \\
   & -128.5475 &      & -32.772443  & 0.157631 & 9.618054 & \\
   &           & $2s$ & -1.930391   & 0.891336 & 1.640769 & $5\times10^{-6}$ \\
   &           &      & -1.930391   & 0.892113 & 1.632553 & \\  
   &           & $2p$ & -0.850410   & 0.967755 & 1.430252 & $6\times10^{-6}$\\
   &           &      & -0.850410   & 0.965274 & 1.435350 & \\
Ar & -526.8030 & $1s$ & -118.610352 & 0.086015 & 17.561606 & $2\times10^{-6}$\\
   & -526.8175 &      & -118.610350 & 0.086104 & 17.553229 & \\
   &           & $2s$ & -12.322153  & 0.411857 & 3.562264 & $2\times10^{-6}$ \\
   &           &      & -12.322153  & 0.412280 & 3.555317 & \\
   &           & $2p$ & -9.571466   & 0.375269 & 3.449283 & $9\times10^{-7}$ \\
   &           &      & -9.571466   & 0.375330 & 3.449989 & \\
   &           & $3s$ & -1.277353   & 1.426944 & 0.967005 & $9\times10^{-5}$ \\
   &           &      & -1.277353   & 1.422172 & 0.961985 & \\
   &           & $3p$ & -0.591017   & 1.668648 & 0.817928 & $5\times10^{-5}$ \\
   &           &      & -0.591017   & 1.662959 & 0.814074 & \\
Kr & -2752.5365 & $1s$ & -520.165467 & 0.042441 & 35.483699 & $5\times10^{-7}$\\
   & -2752.0549 &      & -520.165468 & 0.042441 & 35.498152 &  \\
   &            & $2s$ & -69.903081 & 0.187181 & 7.924967 & $2\times10^{-6}$ \\ 
   &            &      & -69.903082 & 0.187256 & 7.918830 &  \\ 
   &            & $2p$ & -63.009784 & 0.161695 & 7.874355 & $3\times10^{-6}$ \\
   &            &      & -63.009785 & 0.161876 & 7.868429 &  \\ 
   &            & $3s$ & -10.849466 & 0.537875 & 2.644610 & $2\times10^{-6}$ \\ 
   &            &      & -10.849466 & 0.537802 & 2.637556 &  \\
   &            & $3p$ & -8.331501 & 0.542133 & 2.530080 & $2\times10^{-6}$ \\ 
   &            &      & -8.331501 & 0.542627 & 2.522775 &  \\ 
   &            & $3d$ & -3.825234 & 0.550922 & 2.276713 & $4\times10^{-6}$ \\ 
   &            &      & -3.825234 & 0.550880 & 2.276940 &  \\ 
   &            & $4s$ & -1.152935 & 1.630081 & 0.808453 & $1\times10^{-4}$ \\ 
   &            &      & -1.152935 & 1.629391 & 0.804188 &  \\ 
   &            & $4p$ & -0.524186 & 1.950193 & 0.675555 & $3\times10^{-5}$ \\ 
   &            &      & -0.524187 & 1.951611 & 0.669219 &  \\ 
\lasthline
\end{tabular}
\end{table}

Finally, we calculated the total energy for the ground state of 
each atom, by using the following expression:
\begin{equation}
E^{\mathrm{DIM}} = \sum\limits_{nl} 
\left[ 
\varepsilon_{nl}^{\mathrm{DIM}} - 
\frac{1}{2}\int  \rho_{nl}^{\mathrm{DIM}}(r)
\left( V_{nl}^{\mathrm{DIM}}(r) + \frac{Z_{N}}{r}\right) dr \,
\right] \, ,
\label{Eq:Etotal}
\end{equation}
where the density 
$\rho_{nl}^{\mathrm{DIM}}(r) = |u_{nl}^{\mathrm{DIM}}(r)|^2$.
The calculated energies $E^{\mathrm{DIM}}$ are given in 
Table \ref{tab:results}, together with the total energies obtained  
by the Hartree--Fock calculations.
The comparison shows a notable agreement between both 
calculations, at about $0.02\%$.

%%%%%%%%%%%%%%%%%%%%%%%%%%%%%%%%%%%%%%%%%%%%%%%%%%%%%%%%%%%%%%%%%
\subsection{\sffamily \Large The exchange potential}
\label{subsec:exchange}

Orbital--specific exchange potentials can be obtained 
accurately by computing the non--local Fock exchange operator.
A first local approximation can be computed 
with the average exchange charge density proposed 
by Slater\cite{Slater:51}.
Another approximation, proposed by Sharp and 
Horton\cite{Sharp:53}, consists in attaining a 
local potential that approximates the exchange operator 
through a variational procedure that minimizes the energy.
There are several other more elaborated methods that allow 
us to obtain local exchange 
potentials\cite{Krieger:92,Gorling:92,Yang:02,Staroverov:06}.
However, these potentials are rather difficult to put in a 
simple and smooth analytical expression, such as 
Eq.(\ref{eq:atomseq}).

\begin{figure}
\centering
\includegraphics[width=0.4\textwidth]{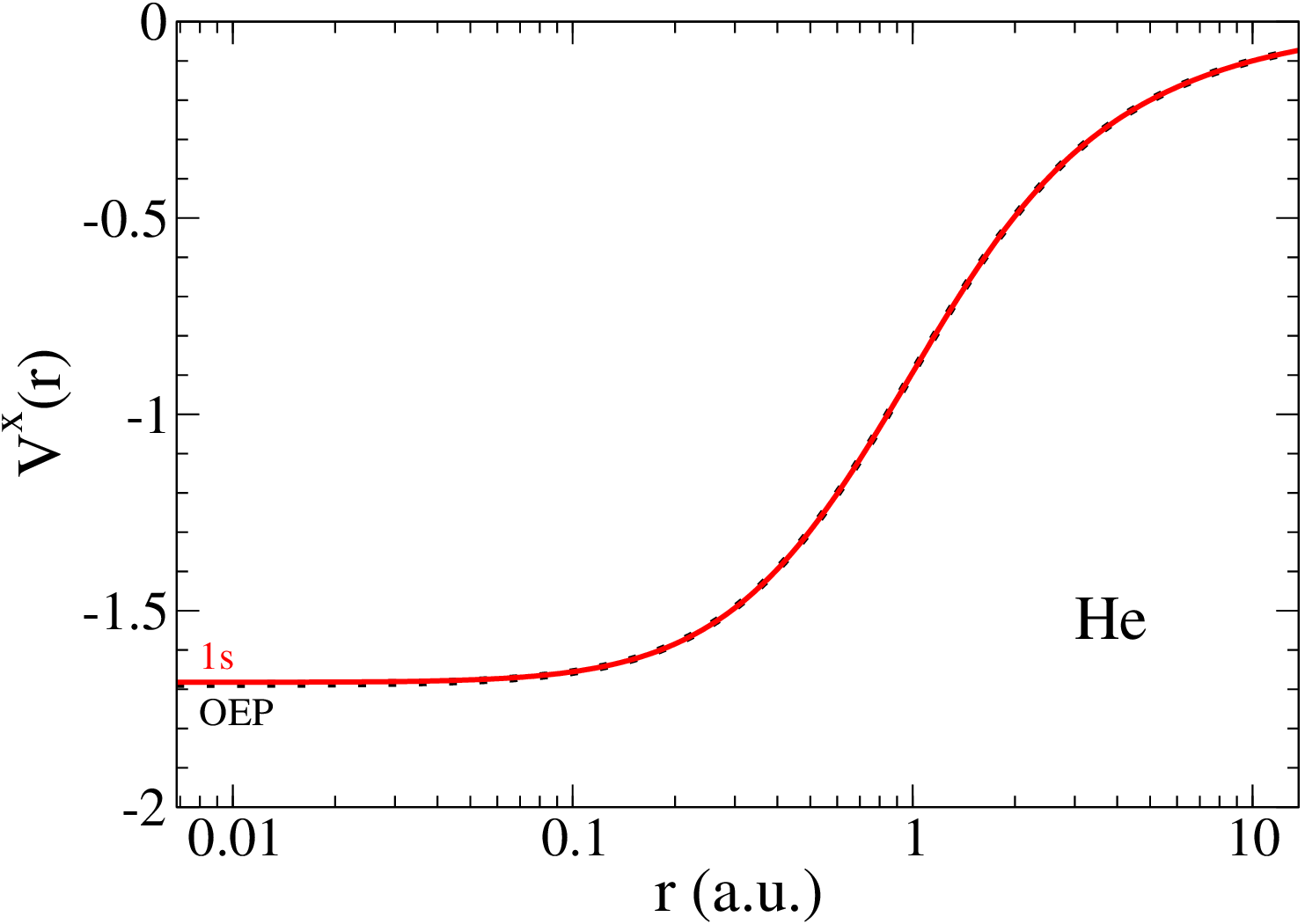} 
\hspace{0.1cm}
\includegraphics[width=0.4\textwidth]{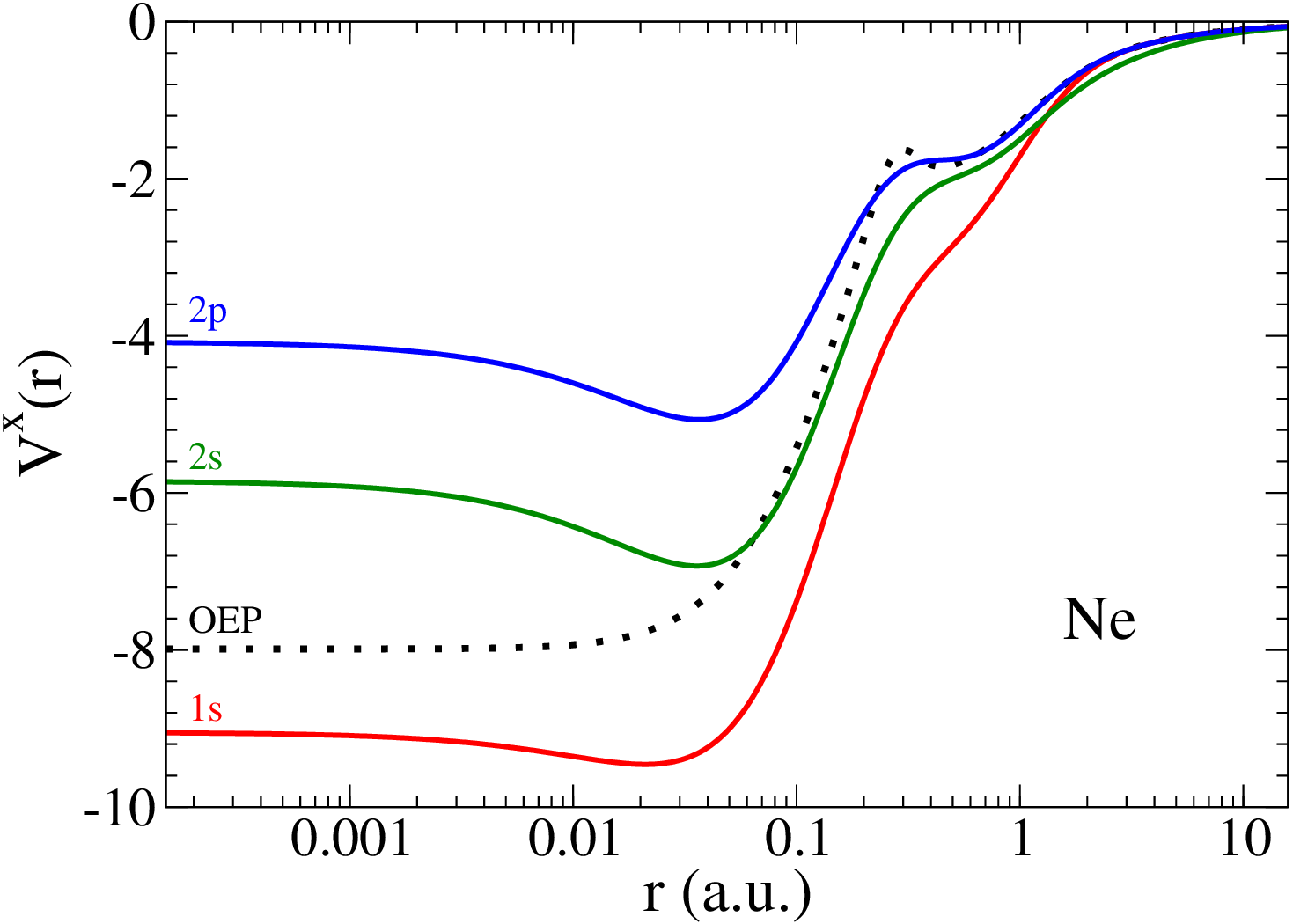} \\
\vspace{0.75cm}
\includegraphics[width=0.4\textwidth]{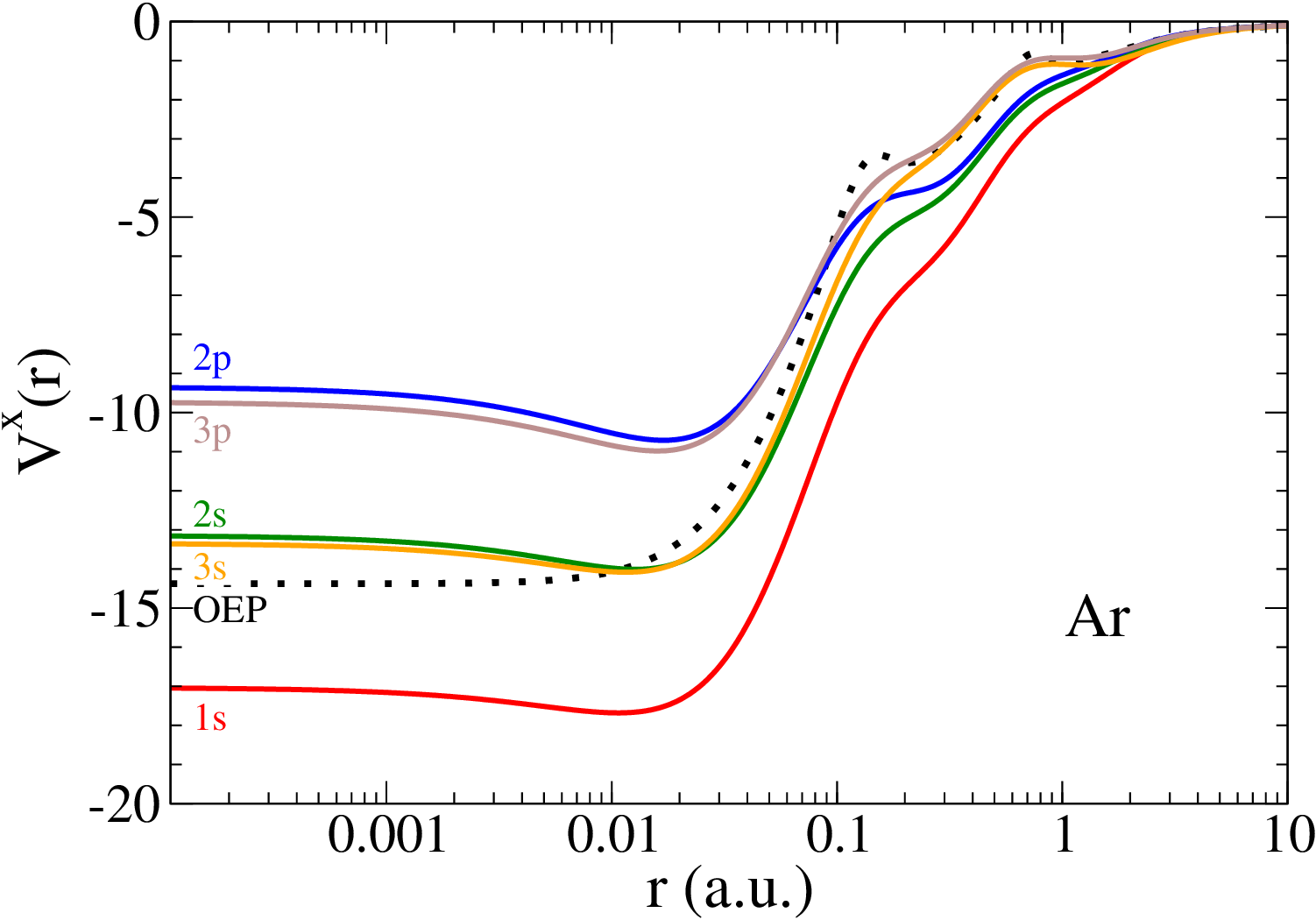}  
\includegraphics[width=0.4\textwidth]{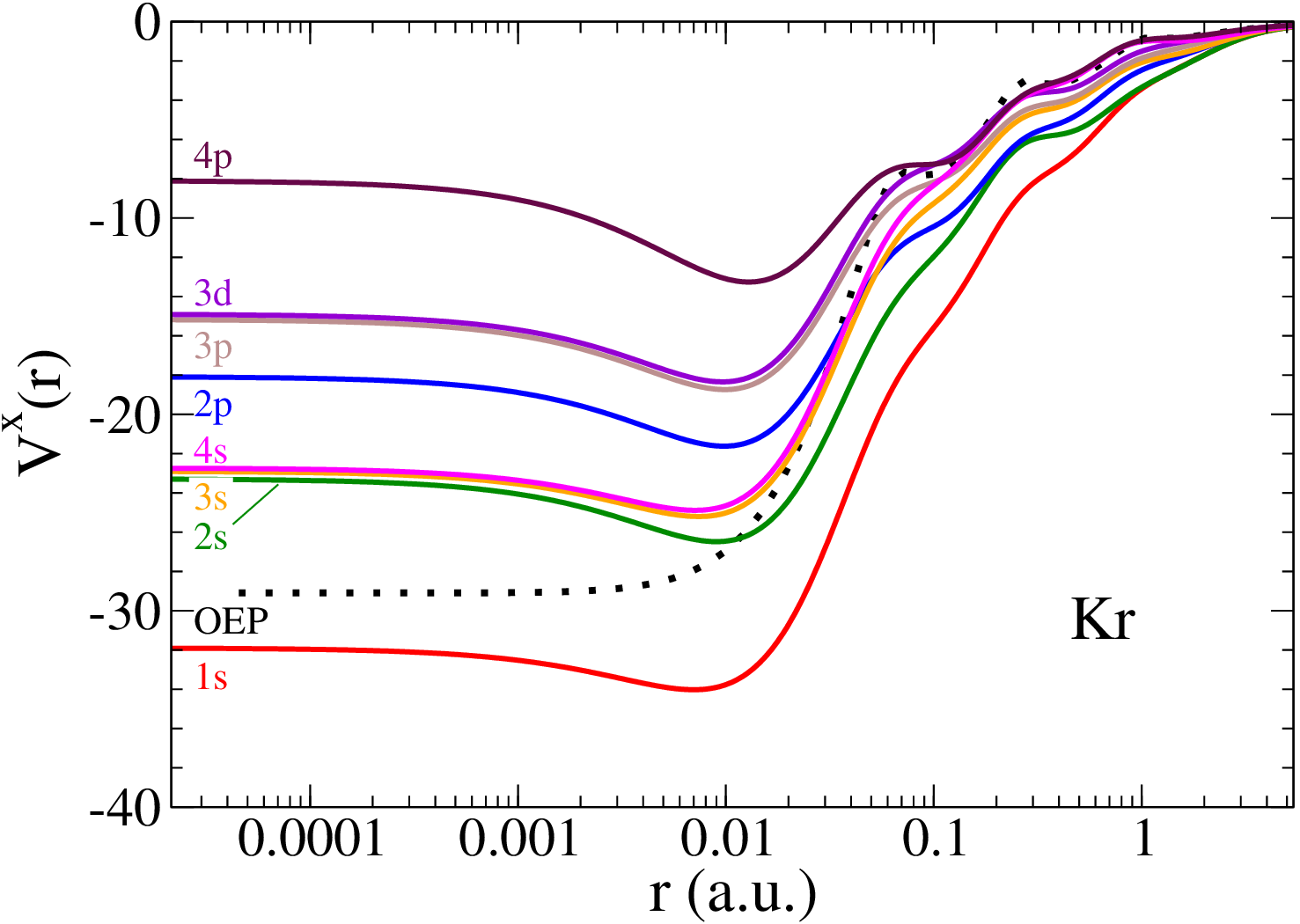}
\caption{(color online). Orbital--specific exchange potentials 
$V_{nl}^{\mathrm{DIMx}}(r)$ and $V^{\mathrm{x}}_{\mathrm{OEP}}$, 
for the ground state of He, Ne, Ar and Kr.}
\label{fig:exchange-HeNeArKr}
\end{figure}

Due to the fact that the Hartree--Fock method does not take 
into account the correlations, our procedure allowed us to obtain 
in a rather direct way ``exact'' local orbital--dependent 
exchange potentials, 
\begin{equation}
V_{nl}^{\mathrm{DIMx}}(r)=V_{nl}^{\mathrm{DIM}}(r)+\frac{Z_{N}}{r}
-\int{ \frac{\rho^{\mathrm{HF}}(r^{\prime})  }
{\left| \mathbf{r} - \mathbf{r^{\prime}} \right|}} \, 
d \mathbf{r^{\prime}} \, ,
\label{eq:exchange-potential}
\end{equation}
where $\rho^{\mathrm{HF}}(r)$ is the total density 
calculated with the $u^{\mathrm{HF}}_{nl}(r)$ wavefunctions. 
Figure \ref{fig:exchange-HeNeArKr} shows
the orbital--specific exchange potentials $V_{nl}^{\mathrm{DIMx}}(r)$ for 
the ground states of the four noble gases He, Ne, Ar, and Kr, 
calculated with the depurated inversion method DIM.

In order to discuss our results, in Fig. \ref{fig:exchange-HeNeArKr}
we plotted the optimized effective potential 
$V_{\mathrm{OEP}}^{\mathrm{x}}(r)$ developed by 
Talman\cite{Talman:89} (black dotted lines) for the noble gases. 
It is well known that the {\sc oep} method finds the potential 
which yields eigenfunctions that minimize the expectation value 
of the Hartree--Fock Hamiltonian.
However, although very accurate, it always 
yields an energy above the HF energy. 
For practical applications the {\sc oep} potential works 
very well for the outer shell.
At longer distances, all the $nl$ orbitals have a 
similar behavior accompanying the {\sc oep} exchange potential.
We noticed that the exchange potentials of the orbitals having 
a common angular momentum $l$ resemble to each other (see Ar
for instance). 
This was suggested in a work by Herman {\it et al.}\cite{Herman:54}
where an $l$--averaged exchange potential for each set of electronic 
states was calculated as a modification of Slater's average 
exchange potential.

According to Eq. (\ref{eq:exchange-potential}) all orbital--specific
potentials should approach the same value at $r=0$, since 
$Z_{nl}^{\mathrm{DIM}}(r)=rV_{nl}^{\mathrm{DIM}}(r)$ approaches $Z_N$
regardless of $nl$ (the second and third term are the same for 
every orbital).
However, from Fig. \ref{fig:exchange-HeNeArKr} it appears
that the potentials for the different orbitals approach different values
at the origin.
This is a consequence of the fact that every DIM potential tends to $Z_N$ 
with different behavior, determined by their fitting parameters. 
In fact, for very low $r$ values $V_{nl}^{\mathrm{DIM}}(r) \approx
\sum_{j}\alpha_{j}\beta_j - V^{\mathrm{d}}(r)$, but they all have 
strictly the same value at $r=0$.

As a final test for our method, we calculated the total exchange 
energy $E^{\mathrm{x}}$ as given by
\begin{equation}
 E^{\mathrm{x}} = \sum_{nl}E_{nl}^{\mathrm{x}} = 
\sum_{nl}\left[\frac{1}{2}\int{\rho^{\mathrm{HF}}_{nl}(r) \, \, 
 V_{nl}^{\mathrm{DIMx}}}(r) \, dr \, \right]
 \label{eq:total-energy}
 \end{equation}
Table \ref{tab:xenergy} displays the orbital exchange energy as well 
as the total exchange energy for He, Ne, Ar and Kr. 
The total exchange energies are compared with the exact atomic 
Hartree--Fock (EAHF) values given by Becke \cite{Becke:88}, with 
very good agreement.
\begin{table}
\caption{ Orbital and total exchange energies 
for He, Ne, Ar, and Kr.}
\label{tab:xenergy}
\centering
\begin{tabular}{c|c|cccc|cc}
\firsthline
%    & \diag{.1em}{.5cm}{$l$}{$n$} & 1 & 2 & 3 & 4 & Total & EAHF \\
   & \backslashbox{$l$}{$n$} & 1 & 2 & 3 & 4 & Total & EAHF \\
  \hline
He &  0  & -0.5129 &  &  &  & -1.0258 & -1.026 \\ 
Ne &  0  & -3.1106 & -0.8620 &  &  &  -12.1080 & -12.11 \\
   &  1  &         & -0.6938 &  &  &  & \\ 
Ar &  0  & -5.8760 & -1.9470 & -0.5742 &  &  -30.1826 & -30.19 \\
   &  1  &         & -1.7974 & -0.4340 &  & \\ 
Kr &  0  &-12.2258 & -4.5523 & -1.9972 & -0.5275 &  -93.8525 & -93.89 \\
   &  1  &         & -4.4305 & -1.8401 & -0.3906 & \\
   &  2  &         &         & -1.5280 &         & \\ 
\hline
\end{tabular}
\end{table}

%%%%%%%%%%%%%%%%%%%%%%%%%%%%%%%%%%%%%%%%%%%%%%%%%%%%%%%%%%%%%%%%%
%%%%%%%%%%%%%%%%%%%%%%%%%%%%%%%%%%%%%%%%%%%%%%%%%%%%%%%%%%%%%%%%%
\subsection{\sffamily \Large Nitrogen DIM and Exchange Potentials}
\label{subsec:Nitrogen}
\begin{figure}
\centering
\includegraphics[width=0.4\textwidth]{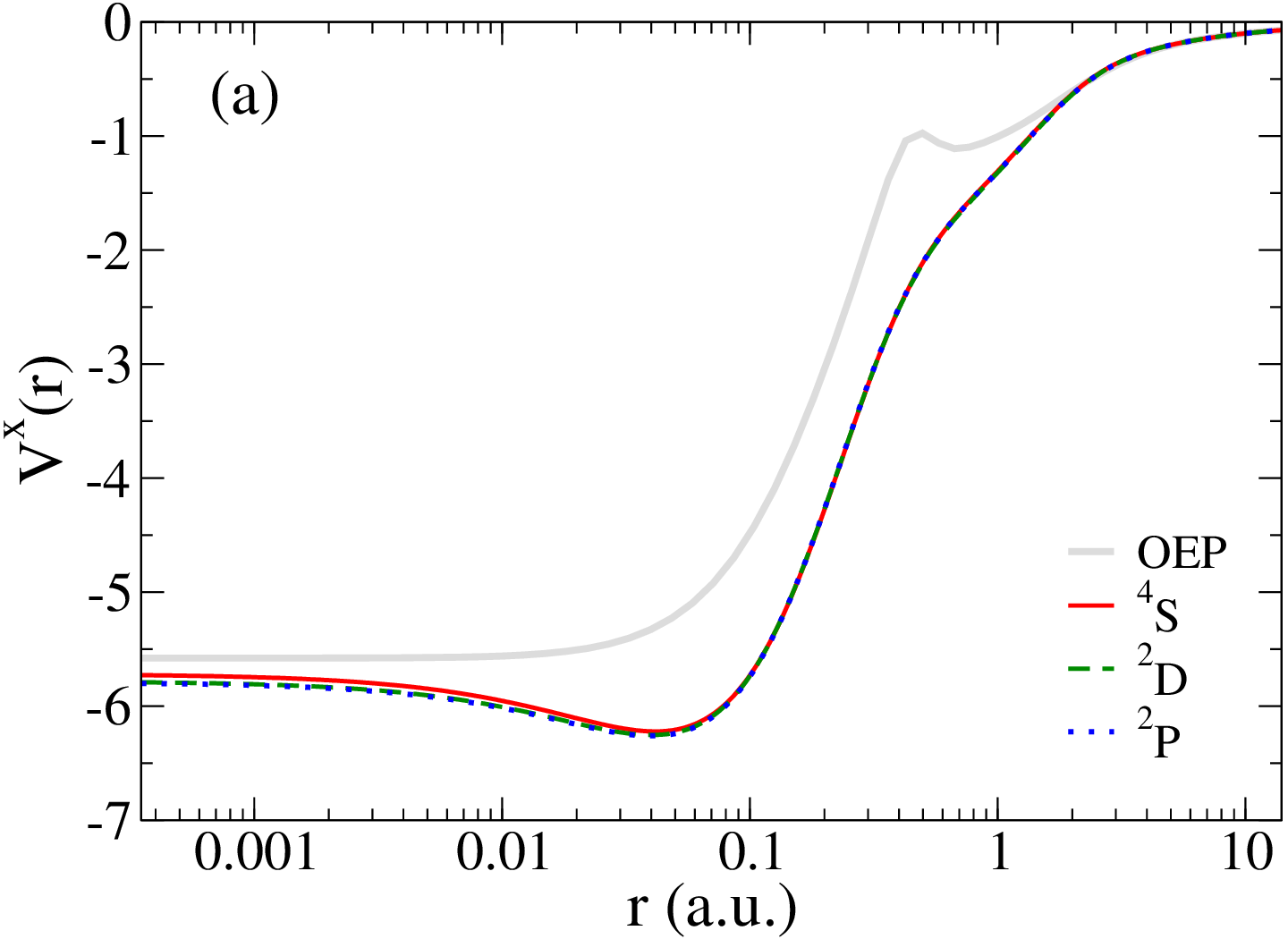} 
\hspace{0.1cm}
\includegraphics[width=0.4\textwidth]{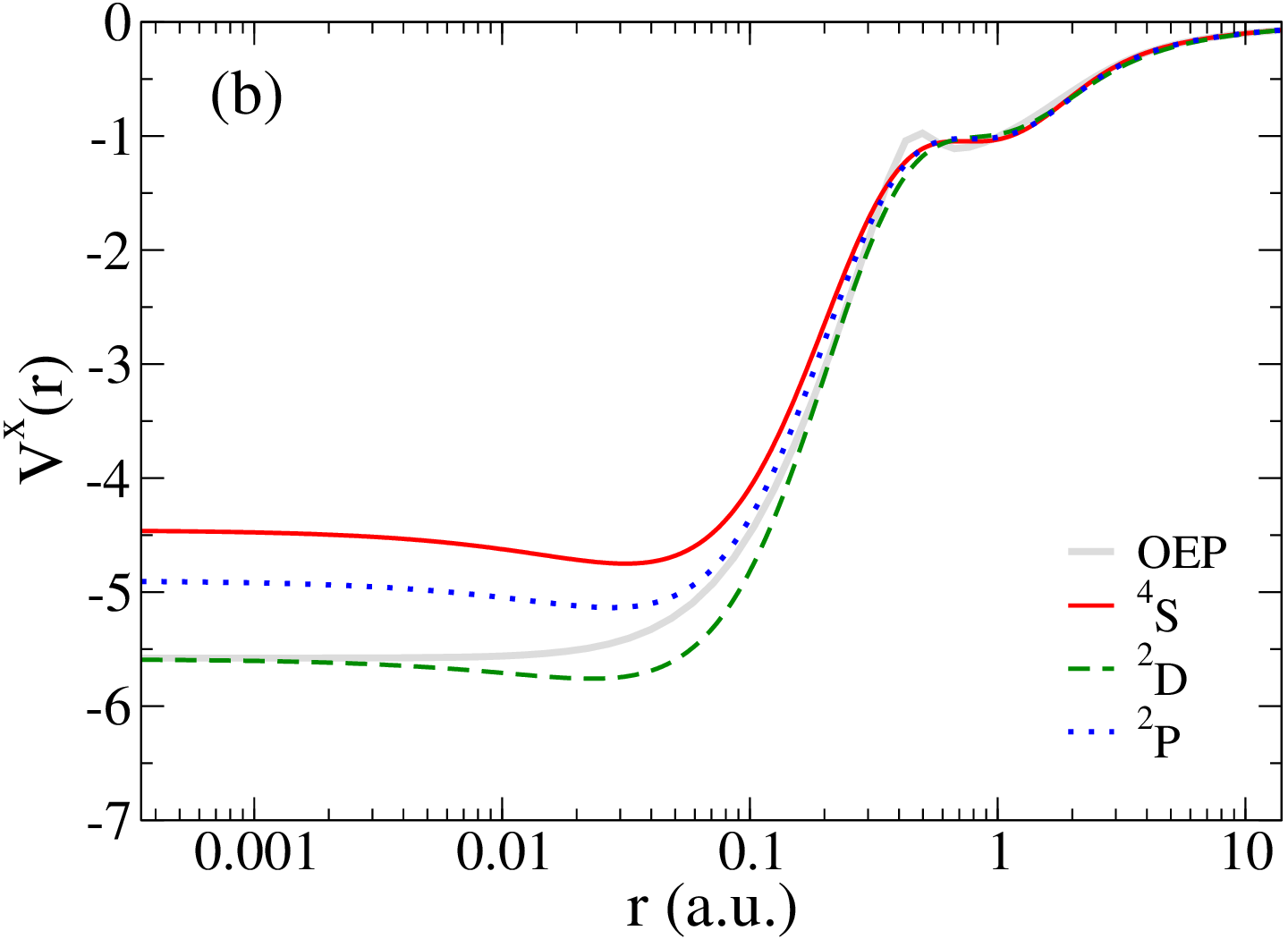} \\
\vspace{0.75cm}
\includegraphics[width=0.4\textwidth]{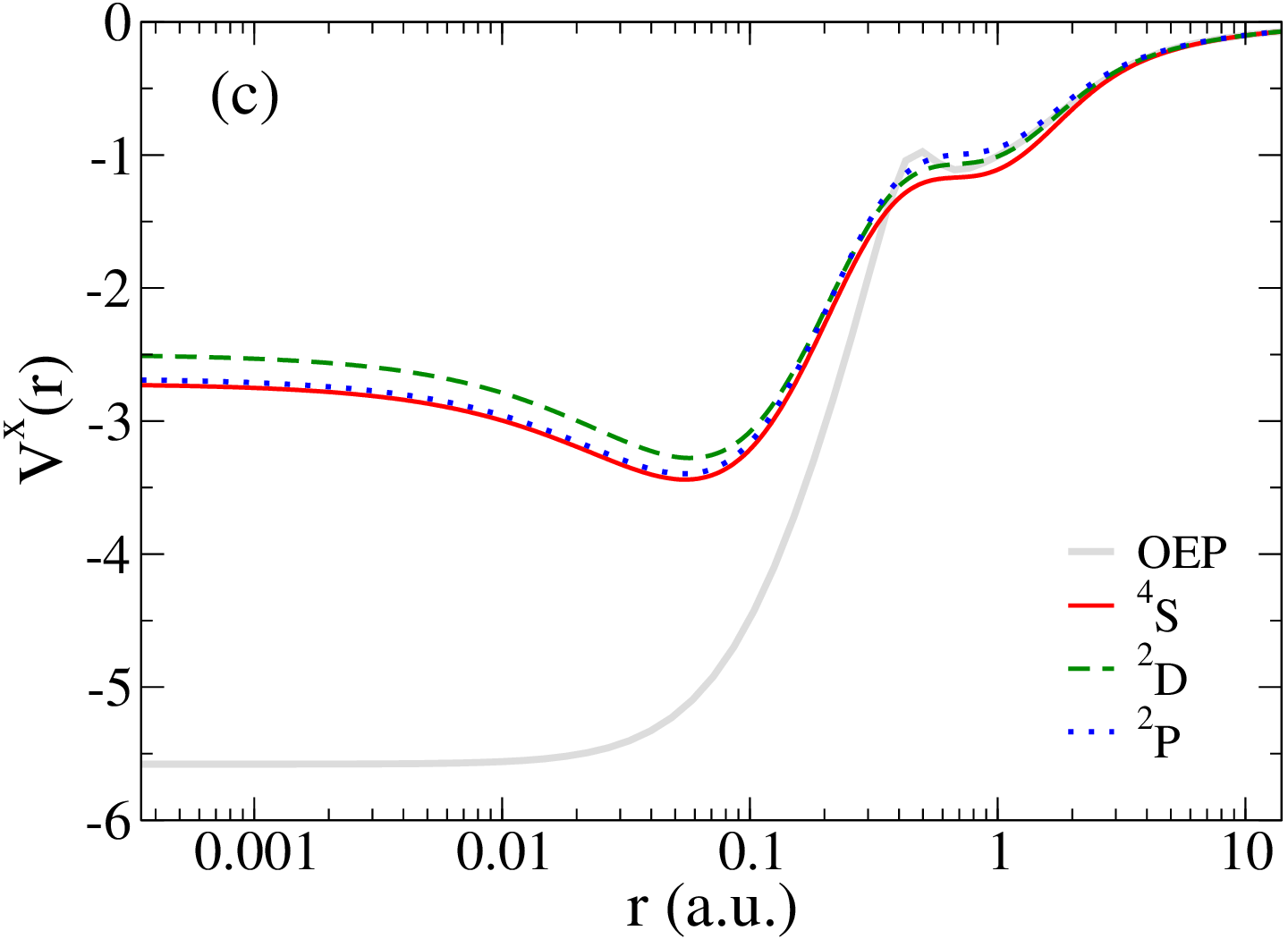}
\caption{(color online). DIM exchange potential 
$V_{nl}^{\mathrm{DIMx}}(r)$ for the (a) $1s$, (b) $2s$ and (c) $2p$ 
orbitals, for the $2\,^4$S, $2\,^2$D and $2\,^2$P terms of Nitrogen.}
\label{fig:nitrogen-exchange}
\end{figure}

The procedure developed here is not limited to noble gases or
closed shells. As an example we will apply the method to Nitrogen. 
The lower configuration $2p^3$ of Nitrogen gives rise to three different 
terms: $2\,^4$S, $2\,^2$D, $2\,^2$P. Each of them is described by 
a different electronic density.
The fitting parameters that define the term--dependent
effective charges are given in Table~\ref{tab:parametersNitro} for
each of the terms. We built the corresponding
DIM potentials from these effective charges.
By using these potentials we solved the Schr\"{o}dinger equation 
(Eq.(\ref{eq:SchrodKS})) for every term, obtained the solutions, 
the energies, and the corresponding mean values 
$\langle r \rangle$ and $\langle 1/r \rangle$. 
The comparison between the orbitals obtained from the diagonalization 
of the Hamiltonian with the effective 
potentials and the original Hartree--Fock orbitals are shown in Table 
\ref{tab:resultsNitro}.
The mean values $\langle r \rangle$ obtained with the DIM effective 
potentials agree with the HF values in about $0.1\%$, and the 
$\langle 1/r \rangle$ mean values agree in about $0.2\%$.
\begin{table}
\caption{Fitting parameters for the effective charge 
$Z_{nl}^{\mathrm{DIM}}(r)$ for $2\,^4$S, $2\,^2$D and $2\,^2$P terms
of Nitrogen.}
\label{tab:parametersNitro}
\centering
\begin{tabular}{ccccccc}
\hline
 & \multicolumn{2}{c}{$2\,^4$S} 
 & \multicolumn{2}{c}{$2\,^2$D } 
 & \multicolumn{2}{c}{$2\,^2$P} \\
% \hline
 $nl$  & $\alpha$ & $\beta$ 
       & $\alpha$ & $\beta$ 
       & $\alpha$ & $\beta$ \\
\hline
 $1s$  & 5.25634  & 1.26207 
       & 5.18635  & 1.22410
       & 5.18635 & 1.21779  \\
\vspace*{0.09cm}
       & 0.743660 & 8.02844 
       & 0.813650 & 7.56800
       & 0.813650 & 7.56740   \\
       %%%%%%%%%%%%%%%%%%%%%%%%%%%%%%%%%
 $2s$  & 2.45281  & 3.51271 
       & 0.398100 & 0.239738
       & 0.890660 & 0.830615  \\
       & 0.833570  & 3.38654 
       & 1.85412  & 1.03105
       & 3.66999  & 3.14946   \\
\vspace*{0.09cm}
       & 2.71362  & 0.894699 
       & 3.74778  & 2.85313
       & 1.43935  & 0.740427  \\
       %%%%%%%%%%%%%%%%%%%%%%%%%%%%%%%%%
 $2p$  & 3.64345  & 1.24069 
       & 4.01052  & 1.28744
       & 1.89769  & 1.16557  \\
       & 2.05501  & 5.35135 
       & 1.85517  & 5.70858 
       & 1.77430   & 5.68782 \\
       & 0.301540 & 0.286609 
       & 0.134310  & 0.267987
       & 2.32801  & 1.40925  \\
\hline
\end{tabular}
\end{table}
The calculated total energies $E^{\mathrm{DIM}}$ for each term
of the Nitrogen atom using Eq.~(\ref{eq:total-energy})
are presented in Table \ref{tab:resultsNitro}.
The agreement between the DIM total energies and the original HF 
total energies is excellent, of about $0.04\%$. 
\begin{table}[htp]
\caption{Total and orbital energy and mean values for the 
$2\,^4$S, $2\,^2$D and $2\,^2$P terms of N obtained from the 
DIM effective potentials (upper rows) compared with the 
Hartree--Fock values (lower rows).} 
\label{tab:resultsNitro}
\centering
\begin{tabular}{c *{1}{S[table-format=3.6]}|c *{1}{S[table-format=5.6]}
*{1}{S[table-format=2.6]} *{1}{S[table-format=2.6]}}
\firsthline
   & {$E$} & {$nl$} &  {$\epsilon$} & {$\langle r \rangle$}
   & {$\langle 1/r \rangle$} \\ 
\hline
$2\,^4$S
 & -54.37617 & $1s$ & -15.62906 & 0.22830 & 6.64863 \\
 & -54.40093 &      & -15.62906 & 0.22830 & 6.65324 \\
 &           & $2s$ & -0.94532  & 1.33448 & 1.08037 \\
 &           &      & -0.94532  & 1.33228 & 1.07818 \\
 &           & $2p$ & -0.56759  & 1.41268 & 0.95498 \\
 &           &      & -0.56759  & 1.40963 & 0.95769 \\
%%%%%%%%%%%%%%%%%%%%%%%%%%%%%%%%%%%%%%%%%%%%%%%%%%%%%%%%%%%%%%
$2\,^2$D 
 & -54.27557 & $1s$ & -15.66639 & 0.22829 & 6.64929 \\
 & -54.29617 &      & -15.66639 & 0.22826 & 6.65388 \\
 &           & $2s$ & -0.96367  & 1.32917 & 1.08644 \\
 &           &      & -0.96367  & 1.32632 & 1.08318 \\
 &           & $2p$ & -0.50866  & 1.44878 & 0.93882 \\
 &           &      & -0.50866  & 1.44662 & 0.94208 \\
%%%%%%%%%%%%%%%%%%%%%%%%%%%%%%%%%%%%%%%%%%%%%%%%%%%%%%%%%%%%%%
$2\,^2$P 
 & -54.20856 & $1s$ & -15.69160 & 0.22824 & 6.65036 \\
 & -54.22810 &      & -15.69160 & 0.22824 & 6.65430  \\
 &           & $2s$ & -0.97634  & 1.32562 & 1.08712 \\
 &           &      & -0.97634  & 1.32232 & 1.08656  \\
 &           & $2p$ & -0.47130  & 1.47176 & 0.92982 \\
 &           &      & -0.47130  & 1.47301 & 0.93155  \\
\hline
\end{tabular}
\end{table}
Figure \ref{fig:nitrogen-exchange} shows the $nl$--orbital exchange 
potentials for the $2\,^4$S, $2\,^2$D and $2\,^2$P terms, 
calculated with the depurated inversion method. 
Again, to compare our results, the exchange potential 
given by Talman\cite{Talman:89} ({\sc oep}) is presented in the 
figures in light grey. 
Figure \ref{fig:nitrogen-exchange}(a) illustrates the 
exchange potential for the $1s$ orbitals for the different terms,
showing an overall similarity. 
The {\sc oep} potential behaves like the
$V_{1s}^{\mathrm{xDIM}}(r)$ only at short and large distances.
Figure \ref{fig:nitrogen-exchange}(b) shows the exchange 
potentials for the $2s$ orbitals. In this case, noticeable
differences between the term--potentials arise at low values of $r$.
For $r$ higher than 0.5 a.u., all the term--potentials become 
indistinguishable and agree perfectly with the {\sc oep} potential.
A pecularity observed in the figure is that the {\sc oep} potential
agrees very well with the $V_{2s}^{\mathrm{xDIM}}(r)$ for the 
$2\,^2D$ term.
Figure \ref{fig:nitrogen-exchange}(c) displays the $2p$ exchange
potentials, which behave similarly for all the terms. However, since 
the {\sc oep} potential is the same for all the orbitals and terms,
it disagrees completely with the $V_{2p}^{\mathrm{x}}(r)$ at short 
distances.

Table \ref{tab:xenergyNitro} presents the total exchange energy
and the $nl$--exchange energies of the $2\,^4$S, $2\,^2$D and 
$2\,^2$P terms.
The $1s$--exchange energy for all the terms are the same, as 
expected for a closed--shell orbital. 
Similarly, the $2s$--exchange energy varies slightly, with a 
difference of $0.08\%$. 
However, this is not the case for the $2p$--exchange energy, which 
varies significantly, having discrepancies of about $18\%$ between the 
different terms. 
The total exchange energy computed with Eq. (\ref{eq:total-energy})
for the terms are compared with the exact atomic Hartree--Fock (EAHF)
exchange energy, with an agreement of about $0.1\%$.
\begin{table}
\caption{ Orbital and total exchange energies for 
$2\,^4$S, $2\,^2$D and $2\,^2$P terms of Nitrogen.}
\label{tab:xenergyNitro}
\centering
\begin{tabular}{cccccc}
\hline
   & 1s & 2s & 2p & Total & EAHF  \\ 
\hline
$2\,^4$S & -2.1175 & -0.4776 & -0.4711 & -6.6034 & -6.596 \\ 
$2\,^2$D & -2.1175 & -0.4777 & -0.4262 & -6.4688 &  \\
$2\,^2$P & -2.1175 & -0.4780 & -0.3973 & -6.3827 &   \\ 
\hline
\end{tabular}
\end{table}

%%%%%%%%%%%%%%%%%%%%%%%%%%%%%%%%%%%%%%%%%%%%%%%%%%%%%%%%%%%%%%%%%
\section{\sffamily \Large CONCLUSIONS}
\label{sec:conclusions}

A crucial requirement of the density functional method is the 
accurate representation of the exchange functional.
On the other hand, the atomic collision community needs accurate 
one--electron potentials in order to generate the bound and 
continuum states on the same footing for further calculations 
of collisional processes. These potentials need to be worked out 
for any $nl$--specific orbital, a feature that in general is 
not present in the chemistry community functionals.
In the present work we devised and implemented a depurated 
inversion method, which allows to obtain the intended potentials 
through a very simple analytical expression of the effective 
charges.
The method consists in the inversion of a Kohn--Sham equation, 
in which the KS orbitals have been replaced by the Hartree--Fock 
orbitals. 
By means of diagonalization we have achieved accurate wavefunctions 
having almost perfect agreement with the original Hartree--Fock 
wave functions.
The quality of the potentials obtained by the present method 
is remarkably good. 
We applied the developed methodology to the calculation of the 
ground state orbitals of noble gases and the Nitrogen 
atom. It is worth mentioning that the same technique can be 
used for any other level, i.e., it is not limited to the 
ground state.

\subsection*{\sffamily \large ACKNOWLEDGMENTS}
This work was supported by grants of CONICET, ANPCyT, and 
UBACyT, of Argentina.

%%%%%%%%%%%%%%%%%%%%%%%%%%%%%%%%%%%%%%%%%%%%%%%%%%%%%%%%%%%%%%%%%%%%%%%
%%%%%%%%%%%%%%%%%%%%%%%%%%%%%%%%%%%%%%%%%%%%%%%%%%%%%%%%%%%%%%%%%%%%%%%
% \section{References}

%%%%%%%%%%%%%%%%%%%%%%%%%%%%%%%%%%%%%%%%%%%%%%%%%%%%%%%%%%%%%%%%%%%%%%%

\end{document}